\newif\ifAnon\Anonfalse
\newif\ifNotes\Notestrue

\newcommand{\swallow}[1]{}
\ifNotes
  \newcommand{\colorcomment}[2]{{\color{#1}[#2]}\xspace}
  
\else
  \newcommand{\colorcomment}[2]{\relax}
  
\fi

\documentclass[conference]{IEEEtran}

\usepackage{xcolor}
\colorlet{Red}{red}
\colorlet{Green}{green}
\usepackage{array}
\usepackage{amssymb}
\usepackage{multirow}
\usepackage[inline]{enumitem}
\usepackage{moreenum}
\usepackage[utf8]{inputenc}
\usepackage{balance}
\usepackage[hidelinks]{hyperref}
\hypersetup{colorlinks=false}
\usepackage{cleveref}
\usepackage{footnote}
\usepackage{tablefootnote}
\usepackage[htt]{hyphenat}
\usepackage[english]{babel}
\usepackage[english]{isodate}
\usepackage{oplotsymbl}
\usepackage{subcaption}
\usepackage[load-configurations=abbreviations,binary-units]{siunitx}
\usepackage{booktabs}
\usepackage{xspace}
\usepackage{url}
\usepackage{pgfplots}
\usepgfplotslibrary{statistics}
\pgfplotsset{compat=1.16}
\usepackage{subfiles}
\usepackage{tikz}
\usepackage{tikz-qtree}
\usepackage{titlesec}

\usetikzlibrary{decorations.pathreplacing,positioning, arrows.meta, trees, quotes, backgrounds}
\usepackage{listings}
\usepackage{stfloats}
\usepackage{wasysym}
\usepackage{comment}
\usepackage{fontawesome5} 
\usepackage{catchfile}
\usepackage{threeparttable}

\usepackage{inconsolata}

\newlist{paraenum}{enumerate*}{1}
\setlist[paraenum]{label=\emph{(\arabic*)}}
\crefname{lstlisting}{listing}{listings}
\Crefname{lstlisting}{Listing}{Listings}

\usepackage{array}
\newcolumntype{L}[1]{>{\raggedright\let\newline\\\arraybackslash\hspace{0pt}}m{#1}}
\newcolumntype{C}[1]{>{\centering\let\newline\\\arraybackslash\hspace{0pt}}m{#1}}
\newcolumntype{R}[1]{>{\raggedleft\let\newline\\\arraybackslash\hspace{0pt}}m{#1}}

\usepackage{threeparttable}

\usepackage{pifont}
\newcommand{\cmark}{\ding{51}}%
\newcommand{\xmark}{\ding{55}}%

\newcommand{\cfbox}[2]{%
    \colorlet{currentcolor}{.}%
    {\color{#1}%
    \fbox{\color{currentcolor}#2}}%
}
\newcommand{\nobox}[1]{%
    \colorlet{currentcolor}{.}%
    {\color{white}%
    \fbox{\color{currentcolor}#1}}%
}

\definecolor{ForestGreen}{RGB}{34,139,34}
\definecolor{BrickRed}{RGB}{199, 0, 57}

\newcommand{\etal}{et~al.\ }
\newcommand{\ie}{\textit{i.e.},\ }

\newcommand{\ddash}{-{}-}

\newcommand{\inc}{\textcolor{BrickRed}{$\blacktriangle$}}
\newcommand{\dec}{\textcolor{ForestGreen}{$\blacktriangledown$}}
\newcommand{\gcmark}{\nobox{\textcolor{ForestGreen}{\cmark}}}
\newcommand{\rxmark}{\nobox{\textcolor{BrickRed}{\xmark}}}

\definecolor{keywords}{HTML}{C54B8C}
\definecolor{types}{HTML}{318CE7}
\definecolor{strings}{HTML}{FAA21A}
\definecolor{symbols}{HTML}{74729A}

\lstdefinestyle{customc}{
    language=C,
    commentstyle=\itshape\color{ForestGreen},
    emph={uint32_t,int16_t,uint8_t,uint8_t,int8_t,size_t,int,char,double,float,unsigned,void,bool},
    emphstyle=\bfseries\color{types},
    keywordstyle=\bfseries\color{keywords},
    stringstyle=\color{strings},
    classoffset=1, 
    otherkeywords={>,<,.,-,!,=,~,|,^,&,+,:},
    morekeywords={>,<,.,-,!,=,~,|,^,&,+,:},
    keywordstyle=\color{symbols},
    classoffset=0,
}

\lstdefinelanguage{llvm}{
	sensitive=true,
	alsoletter={\%},
	comment=[l]{;},
	string=[b]{"},
	keywords=[1]{
add, addrspacecast, alloca, and, ashr, atomicrmw, bitcast, br, call, cmpxchg,
extractelement, extractvalue, fadd, fcmp, fdiv, fence, fmul, fpext, fptosi,
fptoui, fptrunc, frem, fsub, getelementptr, icmp, indirectbr, insertelement,
insertvalue, inttoptr, invoke, landingpad, load, lshr, mul, or, phi, ptrtoint,
resume, ret, sdiv, select, sext, shl, shufflevector, sitofp, srem, store, sub,
switch, to, trunc, udiv, uitofp, unreachable, urem, va_arg, xor, zext
	},
	keywords=[2]{
acq_rel, acquire, addrspace, alias, align, alignstack, alwaysinline, any,
anyregcc, appending, arcp, arm_aapcs_vfpcc, arm_aapcscc, arm_apcscc, asm,
atomic, attributes, available_externally, blockaddress, builtin, byval, c,
catch, cc, ccc, cleanup, cold, coldcc, comdat, common, constant, datalayout,
declare, default, define, dereferenceable, dllexport, dllimport, eq, exact,
exactmatch, extern_weak, external, externally_initialized, false, fast, fastcc,
filter, gc, ghccc, global, hidden, inalloca, inbounds, initialexec, inlinehint,
inreg, intel_ocl_bicc, inteldialect, internal, jumptable, largest, linkonce,
linkonce_odr, localdynamic, localexec, max, min, minsize, module, monotonic,
msp430_intrcc, musttail, naked, nand, ne, nest, ninf, nnan, noalias, nobuiltin,
nocapture, noduplicate, noduplicates, noimplicitfloat, noinline, nonlazybind,
nonnull, noredzone, noreturn, nounwind, nsw, nsz, null, nuw, oeq, oge, ogt, ole,
olt, one, opaque, optnone, optsize, ord, personality, prefix, preserve_allcc,
preserve_mostcc, private, prologue, protected, ptx_device, ptx_kernel, readnone,
readonly, release, returned, returns_twice, samesize, sanitize_address,
sanitize_memory, sanitize_thread, section, seq_cst, sge, sgt, sideeffect,
signext, singlethread, sle, slt, spir_func, spir_kernel, sret, ssp, sspreq,
sspstrong, tail, target, thread_local, triple, true, type, ueq, uge, ugt, ule,
ult, umax, umin, undef, une, unnamed_addr, uno, unordered, unwind, uselistorder,
uselistorder_bb, uwtable, volatile, weak, weak_odr, webkit_jscc, x,
x86_64_sysvcc, x86_64_win64cc, x86_fastcallcc, x86_stdcallcc, x86_thiscallcc,
x86_vectorcallcc, xchg, zeroext, zeroinitializer
	},
	keywords=[3]{
i1, i2, i3, i4, i5, i6, i7, i8, i9, i10, i11, i12, i13, i14, i15, i16, i17, i18,
i19, i20, i21, i22, i23, i24, i25, i26, i27, i28, i29, i30, i31, i32, i33, i34,
i35, i36, i37, i38, i39, i40, i41, i42, i43, i44, i45, i46, i47, i48, i49, i50,
i51, i52, i53, i54, i55, i56, i57, i58, i59, i60, i61, i62, i63, i64, i80, i512,
void, half, float, double, fp128, x86_fp80, ppc_fp128, x86_mmx, label, metadata
	},
}

\lstdefinestyle{llvmir}{
    language=llvm,
    stringstyle=\color{strings},
    commentstyle=\itshape\color{ForestGreen},
    keywordstyle=\color{types},
    keywordstyle=[2]\color{keywords},
	keywordstyle=[3]\color{keywords},
    classoffset=1, 
    otherkeywords={=,\%},
    morekeywords={=,\%},
    keywordstyle=\color{symbols},
    classoffset=0,
}

\begin{document}

\date{}

\title{Fun with flags: How Compilers Break and Fix Constant-Time Code}

\ifAnon
\author{}
\else
\author{\IEEEauthorblockN{Antoine Geimer}
	\IEEEauthorblockA{Univ. Lille, CNRS, Inria \\ Univ. Rennes, CNRS, IRISA\\
		antoine.geimer@inria.fr}
	\and
	\IEEEauthorblockN{Clémentine Maurice}
	\IEEEauthorblockA{Univ. Lille, CNRS, Inria\\
		clementine.maurice@inria.fr}
		}
		
\fi

\maketitle

\begin{abstract}
Developers rely on constant-time programming to prevent timing side-channel attacks. But these efforts can be undone by compilers, whose optimizations may silently reintroduce leaks.
While recent works have measured the extent of such leakage, they leave developers without actionable insights: which optimization passes are responsible, and how to disable them without modifying the compiler remains unclear.

In this paper, we conduct a \textit{qualitative} analysis of how compiler optimizations break constant-time code. 
We construct a dataset of compiler-introduced constant-time violations and analyze the internals of two widely used compilers, GCC and LLVM, to identify the specific optimization passes responsible.  
Our key insight is that a small set of passes are at the root of most leaks.
To the best of our knowledge, we are also the first to characterize how the \textit{interactions} between these passes contribute to leakage.
Based on this analysis, we propose an original and practical mitigation that requires no source code modification or custom compiler: disabling selected optimization passes via compiler flags.
We show that this approach significantly reduces leakage with minimal performance overhead, offering an immediately deployable defense for developers.
\end{abstract}

\section{Introduction}

The constant-time programming discipline has been widely applied to cryptographic libraries to protect code against side-channel vulnerabilities.
Writing correct constant-time code can however be difficult: besides avoiding secret-dependent branches and memory accesses at source code level, the programmer must also make sure this property is preserved by the compilation step.
The research community has addressed this issue by creating programming languages dedicated to cryptography~\cite{AlmeidaBB17,CauligiSJ19}, as well as verified constant-time preserving compilers such as CompCert~\cite{BartheBG20}.
However, using the former requires rewriting large parts of cryptographic libraries in a new language, while integrating the latter in an existing build system might be difficult.
Unfortunately, we note that in practice commonly-used libraries and distributions do not make use of these contributions. 
Instead they use common compilers such as the GCC or LLVM, which do not include any ways to mark data as secret, nor do they consider side channels in their models.
As such, it has been known for a while that such compilers can produce vulnerable binaries from constant-time source code~\cite{SimonCA18,SchneiderLP24}.

This makes writing side-channel secure cross-platform code an exceedingly arduous task, leading to programmers adopting two different approaches to solve this issue.
On one hand, they can re-implement critical functions in assembly snippets for each targeted architecture -- a time-consuming task that risk introducing more bugs.
On the other hand they can purposefully complexify their code to counter the compiler's optimizations -- hardly a resilient approach as compilers improve.

\textbf{Problem.} While a mix of both approaches is generally applied in cryptographic libraries, compiler-introduced side-channel vulnerabilities are still regularly found~\cite{BernsteinBB25,PurnalKyber24}.
In fact, recent studies showed that such vulnerabilities might be much more common than previously thought~\cite{SchneiderLP24,GerlachPS25}.
Newer compiler versions include optimizations able to \textit{undo} most of the added complexity, reverting it to a constant-time violation.
One approach would be to actively patch existing compilers to make them side-channel aware.
This requires a dual effort: an engineering one to get the patch working first, and a social one to get it accepted and upstreamed by maintainers who often have little bandwidth for additional worries.
In fact, previous efforts to introduce a constant-time builtin select function in LLVM have remained unmaintained~\cite{ctchoose}, and although many were generally favorable to the idea~\cite{LLVMdev19}, the preferred approach was summarized as: \textit{``I think if you want to force a particular instruction to be used, there is already a pretty reasonable approach: inline assembly''}.
Similar feature requests were made in GCC, similarly to no avail~\cite{GCCMailCmov}.
However, relying on assembly snippets might not be such a safe approach in the future, as we have no guarantees that future compiler versions will not attempt to optimize them.
In fact, it was noted that the SUNWspro compiler would treat inline assembly as templates and optimize them~\cite{Pornin25}.
This overall depicts a rather worrying picture of side-channel security in compilers: current countermeasures are failing, while the inertia behind large software systems means that more principled approaches are unlikely to be used.

\textbf{Goal.} Our paper aims at restoring some trust in off-the-shelf compilers, by providing a clear picture of \textit{how} they break constant-time and how we can prevent it.
While a previous study~\cite{SchneiderLP24} provides an extensive quantification of how often such vulnerabilities occur, only a few details are given as to which compiler passes introduce them or how they interact with each other.
Additionally, it remains unclear if it is possible to keep using these compilers while preventing problematic optimizations from being applied.
By first assembling a dataset of compiler-introduced constant-time violations, we precisely determine the interactions between optimization passes that introduced them. We can then deduce a set of optimizations to disable using option flags, and evaluate how effective this approach is.

\noindent\fbox{%
\begin{minipage}{\linewidth-.3cm}
\textbf{Requirements.} Our approach is guided by practical constraints that reflect the workflows of real-world developers, as emphasized in prior studies~\cite{JancarFB22,FourneBJSSBFA24}. Specifically, we adopt the following two requirements:
\begin{description}
  \item[R1] Avoid requiring any source code changes or annotations;
  \item[R2] Avoid modifying the compiler or relying on non-standard toolchains.
\end{description}
These constraints rule out many existing approaches, such as using verified compilers or constant-time languages, which require either significant rewriting efforts or custom infrastructure. Instead, we focus on improving the security of code compiled with unmodified, widely-used compilers like GCC and LLVM.
\end{minipage}
}
\vspace{.2cm}

We tackle the following research questions: 
\begin{description}
\item[RQ1] How can we automatically detect compiler-introduced constant-time bugs?
\item[RQ2] Which set of compiler optimizations can introduce constant-time bugs and how?  
\item[RQ3] Can we prevent current compilers from introducing constant-time bugs while preserving performance? 
\end{description}

\textbf{Contributions} We make the following contributions:
\begin{enumerate}
\item We provide a simple methodology to discover compiler-introduced constant-time bugs using Microwalk~\cite{WichelmannME18} (\Cref{sec:dataset}).
We analyze the internal behavior of compilers to understand which specific optimization passes break constant-time source code and how they interact. In contrast to related work that typically focuses on optimization levels or compiler versions, we offer a finer-grained analysis at the level of individual passes (\Cref{sec:analysis}).
\item We present a practical and \emph{immediately actionable} mitigation, based on disabling selected (and sometimes undocumented) optimization passes via compiler flags. Our approach applies to both GCC and LLVM and significantly reduces compiler-induced leakage (\Cref{sec:evaluation:CT}).
\item We evaluate the effectiveness and performance impact of our mitigation strategy. We show that disabling a small number of passes removes all compiler-introduced vulnerabilities resulting from constant-time source code, with low overhead (\Cref{sec:evaluation:perf}).
\end{enumerate}

The anonymized artifact can be found at here: \url{https://zenodo.org/records/15207769}.

\section{Background}\label{sec:background}

\subsection{Side-channel Vulnerabilities}

Side-channel attacks focus on exploiting information leakage from side-effects in software execution. There is an extensive literature on attacks exploiting physical side channels such as power consumption~\cite{KocherJJ99}, and electromagnetic emanations~\cite{AgrawalARR02}, as well as microarchitectural side channels~\cite{LouZJ21}. Microarchitectural side channels can be exploited purely with software: the attacker needs no \textit{physical access}, merely execution rights on the device.
The attacker can thus deduce the victim's activity by using a spy program to measure the state of a wide variety of microarchitectural components such as caches~\cite{Percival05,Bernstein05,YaromF14}, branch prediction units~\cite{AciicmezKS07}, execution ports~\cite{AldayaBu19}, prefetchers~\cite{ShinKK18} and many others.
In practice, any microarchitectural component that is \textit{shared} between programs can be an attack vector. In conjunction, any program whose execution induces different microarchitectural states for different secret values is vulnerable.

\subsection{Constant-time Programming}

To mitigate against such attacks, cryptographic developers must follow constant-time (CT) programming principles~\cite{IntelCTGuidelines}.
Code is required to be designed such that the microarchitectural state of the system remains independent of any secret-derived value.
CT is commonly implemented by avoiding any secret-dependent branches and memory accesses. The former prevents timing and branch prediction attacks while the latter prevents cache attacks.
Another often considered requirement is avoiding computations on secret values using instructions whose timing is operand-dependent, such as multiplication~\cite{GrossschadlOP10} or division~\cite{BernsteinBB25} in some architectures.

\begin{figure}[t]
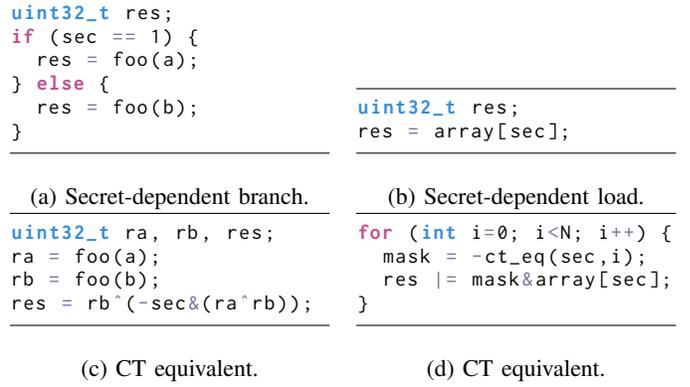

\begin{subfigure}[b]{0.48\columnwidth}
    \begin{lstlisting}
uint32_t res;
if (sec == 1) {
  res = foo(a);
} else {
  res = foo(b);
}
    \end{lstlisting}
    \caption{Secret-dependent branch.}
\end{subfigure}%
    \hfill
\begin{subfigure}[b]{0.48\columnwidth}
    \begin{lstlisting}
uint32_t res;
res = array[sec];
    \end{lstlisting}
    \caption{Secret-dependent load.}
\end{subfigure}%
\\
\begin{subfigure}{0.48\columnwidth}
    \begin{lstlisting}
uint32_t ra, rb, res;
ra = foo(a);
rb = foo(b);
res = rb^(-sec&(ra^rb));
    \end{lstlisting}
    \caption{CT equivalent.}
\end{subfigure}
    \hfill
\begin{subfigure}{0.48\columnwidth}
    \begin{lstlisting}
for (int i=0; i<N; i++) {
  mask = -ct_eq(sec,i);
  res |= mask&array[sec];
}
    \end{lstlisting}
    \caption{CT equivalent.}
    \label{fig:ct-patterns-4}
\end{subfigure}
\caption{Secret-dependent control-flow and memory accesses mitigated through CT programming.}
\label{fig:ct-patterns}
\end{figure}

In practice, secret-dependent control flows are eliminated through \textit{linearization}: both sides of the conditional are run, and the right result is selected using bit-mask arithmetic. Secret-dependent memory accesses are eliminated similarly, by accessing all elements of an array and selecting the right result using a mask. \Cref{fig:ct-patterns} illustrates how both cases of side-channel leakages are solved using CT programming, these patterns being extremely common in cryptographic libraries. 

\lstset{style=customc}

\subsection{Compiler Optimizations}

To ease the implementation of new languages and architectures, compiler suites are usually split into three parts. The front-end is language-specific, and handles the translation from source code to an intermediate representation (IR).
The back-end is architecture-specific, and lowers the IR into assembly, applying some architecture-specific optimizations.
The middle-end optimizes the IR by applying a number of platform-agnostic optimization passes. 
Clang is based on LLVM and thus uses LLVM IR, while GCC's middle-end employs multiple IRs depending on the compilation stage. 

We detail here a few optimization passes in LLVM and GCC.

\textbf{Loop unswitching [LLVM/GCC].} Loop unswitching~\cite{LoopUnswitchingLLVM,GCCOptimizations} is applied to loops containing conditionals independent of the loop's body.
As conditional jumps can be expensive, especially if mispredicted, this optimization brings the conditional outside the loop's body, instead duplicating the loop in each side of the branch.
As this operation can double the overall size of loop, it is only applied when deemed profitable by a heuristic.

\textbf{Vectorization [LLVM/GCC].} Vectorization~\cite{VectorizationLLVM} is a common technique to improve performance of parallelizable code by using SIMD instructions.
Consecutive loop iterations manipulating scalars can thus be merged into a sequence of vector operations.
For example, a loop iterating over an array of 64-bit integers can be parallelized by using instructions on 256-bit vectors.
In this case, 4 integers can be processed in a single iteration.
Additional logic is necessary in case the loop bound is not a multiple of the vectorization factor, or to check if the arrays might overlap each others.
As LLVM IR supports vector instructions, vectorization is done in LLVM's middle-end, with the back-end lowering the IR into appropriate SIMD instructions if available.

\textbf{\texttt{cmov} conversion [LLVM].} \texttt{cmov} conversion~\cite{CmovConversionLLVM} is an optimization present in the LLVM x86 back-end which replaces conditional move instructions with conditional jumps, if deemed profitable. The conversion is deemed profitable if it is done within an innermost loop's body and if the added branch is predicted well enough. Similar optimizations also exist in LLVM's middle-end, replacing \texttt{select} IR instruction by conditional jumps.

\textbf{Jump threading [GCC].} Jump threading~\cite{GCCOptimizations} is an optimization that reduces the number of jumps by merging jumps with partial overlapping conditions or removing them altogether. While this optimization is also present in LLVM, its impact in GCC is more relevant to our study, due to the tendency of GIMPLE IR to represent the ternary conditional operator as a branch by default.

\textbf{Path splitting [GCC].} 
Similarly, path splitting~\cite{GCCOptimizations} is applied in loops containing if-then-else structures. This transformation duplicates the loop body's exit into both sides of the branch in order to save an unconditional jump.

\section{Finding Compiler-introduced Bugs}\label{sec:dataset}

In this section, we detail our methodology to find compiler-introduced bugs (\textbf{RQ1}). We start by highlighting the challenges, and then our setup to address them. 

\subsection{Challenges}

\textbf{C1: Ground truth.} Detecting compiler-introduced constant-time violations is challenging, as it represents a two-fold problem.
First, it requires accurate detection of constant-time violations at binary-level.
Second, we must make sure that such violations are not already present in the source code.
Unfortunately, designing an accurate constant-time analysis that remains scalable for both source code and binary is still an open research question~\cite{GeimerVR23}.

One way to deal with this issue is to base the analysis on binaries whose source code have been verified to be constant-time, such as HACL\*~\cite{DBLP:conf/ccs/ZinzindohoueBPB17}, or on a set of toy examples~\cite{SimonCA18}.
However this limits the scope of the study, as many cryptographic libraries used by developers in practice are not verified.
Another approach can be to include unverified libraries and heuristically filter out non constant-time functions. Previous works have adopted this approach by manually keeping a list of filtered out functions~\cite{SchneiderLP24}.
As they noted however, such manual curation can be incomplete, and function inlining can still bring filtered-out functions into the analysis' scope. 
As such, previous works suffer from a lack of clear ground truth.

\textbf{C2: Comparing binaries.} Instead of filtering out non-constant-time functions, we propose to find compiler-introduced constant-time violations by compiling the same benchmark under different compiler versions.
Previous works have noticed that more recent compiler versions tend to break constant-time guarantees more often~\cite{SchneiderLP24,SimonCA18}.
We base our analysis on the insight that if the same source code is compiled using an older version and a more recent version of LLVM, the added vulnerabilities should be caused by the compiler.
This allows us to side-step \textbf{C1} while providing a more precise analysis. 

One issue with this approach is that binary-level constant-time detection tools often report violations at \textit{instruction level}.
With such metrics, compiler optimizations such as function inlining or loop unrolling can artificially inflate the number of violations found.
As the binaries produced will be different, the same violation can also be identified by a different address, making it impossible to clearly identify if a given constant-time violation was added by a compiler or not.
To design a meaningful analysis, we need to find a way to easily \textit{compare different binaries}.

\subsection{Setup} \label{sec:setup}

We now detail the system parameters for our study, the compilers and source code chosen, as well as the vulnerability detection tool chosen.

\textbf{System.} We pick an older and a newer version of both GCC and LLVM to compare the number of constant-time violations produced.
For LLVM, we pick version~12 and version~18, as previous works noted that LLVM versions following version 12 were more likely to produce non-CT binaries~\cite{SchneiderLP24}.
For GCC, we pick version~9 and version~13.
We compare binaries obtained using optimizations levels \texttt{O3} and \texttt{Os} as they are known to introduce constant-time violations.
All experiments in the remainder of this paper are run on an HP EliteBook 840 G8 laptop, equipped with an Intel~i7-1185G7 processor and running Ubuntu~24.04.2~LTS (kernel version 6.8).
    
\textbf{Benchmarks.}
While we do not need to completely address \textbf{C1}, we still assemble a benchmark with a clearer ground truth to ease the interpretation of our results.
We reuse a subset of the benchmark introduced in \cite{GeimerVR23}, which includes cryptographic functions from BearSSL~\cite{web:bearssl} and MbedTLS~\cite{web:mbedtls}, adapting it to support multiple compilers.
Thanks to the multiple detection tools they used, we can be reasonably certain of which benchmark is constant-time.
In addition, we add some constant-time toy example functions that have been known to be broken by compilers in the past.
This includes the \texttt{ct\_select} functions used in Simon \etal~\cite{SimonCA18}, \texttt{poly\_frommsg} from Kyber which was recently found to be broken by LLVM~\cite{PurnalKyber24} and other examples that were reported on GCC's bug tracker~\cite{GCCJumpThreadingBug,GCCPathSplittingBug}.
To better reproduce the conditions in which constant-time violations were noted, we compile these toy examples specifically with loop unrolling and vectorization disabled.
The code used for our toy examples can be seen in the Appendix.

\textbf{Vulnerability detection.}
To precisely differentiate compiler-introduced constant-time violations from those stemming from non-CT source code, we compare the number of violations found in binaries generated using different compiler versions.
We analyze these binaries with the constant-time vulnerability detection tool Microwalk~\cite{WichelmannME18}.
Microwalk uses the Intel \texttt{Pin} dynamic instrumentation framework to collect multiple execution traces of the binary under different secret inputs, then compares them.
The choice of detection tool is significant for our study.
As a dynamic analysis tool, Microwalk can miss leakages not covered by the traces and in some case, vulnerabilities found can ``obscure'' others, leading to false negatives.
Using a sound static analysis tool could have prevented such false negatives.
However our experiment would have suffered from the lack of scalability of such tools (we compile in total 152 binaries), and static analysis conversely often leads to false positives. 
Finally, Microwalk benefits from active development, being recently extended to improve its usability among developers~\cite{WichelmannSP22,FayolleWK24}.

\textbf{Comparison metric.}
Microwalk reports vulnerable instructions by their addresses.
One approach to compare these results could be to count the number of vulnerable instructions. However this does not give a reliable picture of constant-time violations added by later compiler versions.
Broader application of function inlining and loop unrolling tends to duplicate vulnerable instructions, leading to higher counts.
Even accounting for this, such metric does not allow us to properly compare results from one compiler version to another, as the addresses used in the binaries will be different.
To properly solve \textbf{C2}, we use DWARF debug symbols (with \texttt{-g}) to associate each vulnerable instruction with the source code line it originates from.
This allows us to determine for each vulnerable instruction reported with one compiler version if it can also be found with another version, or if it was added by the former.
Details on the impact of loop unrolling on our experiments and how our metric alleviates it can be found in the Appendix.

\subsection{Experiments and Results}

\newcommand{\tablecompilersdiff}{AES-CBC-bearssl (T) & 32 & 32 & 32 & 32 & 25 & 25 & 25 & 25 \\
AES-GCM-bearssl (T) & 32 & 32 & 32 & 32 & 25 & 25 & 25 & 25 \\
AES-CBC-mbedtls (T) & 23 & 23 & 23 & 23 & 23 & 23 & 22 & 22 \\
AES-GCM-mbedtls (T) & 26 & 26 & 26 & 26 & 26 & 26 & 25 & 25 \\
RSA-mbedtls (PKCS) & 47 & 47 & 51 & \textbf{50} \dec & 52 & \textbf{48} \dec & 50 & \textbf{54} \inc \\
RSA-mbedtls (OAEP) & 46 & \textbf{48} \inc & 50 & \textbf{49} \dec & 49 & 49 & 50 & 50 \\
ECDSA-mbedtls & 60 & \textbf{64} \inc & 63 & \textbf{65} \inc & 61 & \textbf{62} \inc & 62 & \textbf{69} \inc \\
\midrule
AES-CBC-bearssl (BS) & 0 & 0 & 0 & 0 & 0 & 0 & 0 & 0 \\
AES-GCM-bearssl (BS) & 0 & 0 & 0 & 0 & 0 & 0 & 0 & 0 \\
PolyChacha-bearssl (CT) & 0 & 0 & 0 & 0 & 0 & 0 & 0 & 0 \\
PolyChacha-bearssl (SSE2) & 0 & 0 & 0 & 0 & 0 & 0 & 0 & 0 \\
RSA-bearssl (OAEP) & 0 & \textbf{1} \inc & 0 & \textbf{1} \inc & 0 & 0 & 0 & 0 \\
ECDSA-bearssl & 0 & \textbf{1} \inc & 0 & \textbf{1} \inc & 0 & 0 & 0 & 0 \\
PolyChacha-mbedtls & 0 & 0 & 0 & 0 & 0 & 0 & 0 & 0 \\
ct\_select$^\dag$ & 0 & 0 & 0 & 0 & 0 & 0 & 0 & 0 \\
poly\_frommsg$^\dag$ & 0 & \textbf{1} \inc & 0 & \textbf{1} \inc & 0 & 0 & 0 & 0 \\
jump\_threading$^\dag$ & 0 & 0 & 0 & 0 & 1 & 1 & 1 & 1 \\
loop\_unswitching$^\dag$ & 1 & 1 & 0 & 0 & 1 & 1 & 0 & 0 \\
path\_splitting$^\dag$ & 0 & 0 & 0 & 0 & 1 & 1 & 0 & 0 \\
}
\begin{table*}[ht]
\centering
\caption{Number of vulnerable source code lines per binaries, comparison between LLVM 12 and LLVM 18, GCC 9 and GCC 13 for the options \texttt{O3} and \texttt{Os}. We highlight increases in number of vulnerabilities with \inc{} and decreases with \dec.  The first group of binaires is produced from non-CT source code, and the second one from CT source code.}
\label{tab:compilersdiff}
\small
\begin{threeparttable}
\begin{tabular}{l|cccc|cccc}
  \toprule
     & \multicolumn{4}{c}{LLVM} & \multicolumn{4}{c}{GCC} \\
     & \multicolumn{2}{c}{O3} & \multicolumn{2}{c}{Os} & \multicolumn{2}{c}{O3} & \multicolumn{2}{c}{Os} \\
  
  Binaries & LLVM 12 & LLVM 18 & LLVM 12 & LLVM 18 & GCC 9 & GCC 13  & GCC 9 & GCC 13 \\

  \midrule

  AES-CBC-bearssl (T) & 32 & 32 & 32 & 32 & 25 & 25 & 25 & 25 \\
AES-GCM-bearssl (T) & 32 & 32 & 32 & 32 & 25 & 25 & 25 & 25 \\
AES-CBC-mbedtls (T) & 23 & 23 & 23 & 23 & 23 & 23 & 22 & 22 \\
AES-GCM-mbedtls (T) & 26 & 26 & 26 & 26 & 26 & 26 & 25 & 25 \\
RSA-mbedtls (PKCS) & 47 & 47 & 51 & \textbf{50} \dec & 52 & \textbf{48} \dec & 50 & \textbf{54} \inc \\
RSA-mbedtls (OAEP) & 46 & \textbf{48} \inc & 50 & \textbf{49} \dec & 49 & 49 & 50 & 50 \\
ECDSA-mbedtls & 60 & \textbf{64} \inc & 63 & \textbf{65} \inc & 61 & \textbf{62} \inc & 62 & \textbf{69} \inc \\
\midrule
AES-CBC-bearssl (BS) & 0 & 0 & 0 & 0 & 0 & 0 & 0 & 0 \\
AES-GCM-bearssl (BS) & 0 & 0 & 0 & 0 & 0 & 0 & 0 & 0 \\
PolyChacha-bearssl (CT) & 0 & 0 & 0 & 0 & 0 & 0 & 0 & 0 \\
PolyChacha-bearssl (SSE2) & 0 & 0 & 0 & 0 & 0 & 0 & 0 & 0 \\
RSA-bearssl (OAEP) & 0 & \textbf{1} \inc & 0 & \textbf{1} \inc & 0 & 0 & 0 & 0 \\
ECDSA-bearssl & 0 & \textbf{1} \inc & 0 & \textbf{1} \inc & 0 & 0 & 0 & 0 \\
PolyChacha-mbedtls & 0 & 0 & 0 & 0 & 0 & 0 & 0 & 0 \\
ct\_select$^\dag$ & 0 & 0 & 0 & 0 & 0 & 0 & 0 & 0 \\
poly\_frommsg$^\dag$ & 0 & \textbf{1} \inc & 0 & \textbf{1} \inc & 0 & 0 & 0 & 0 \\
jump\_threading$^\dag$ & 0 & 0 & 0 & 0 & 1 & 1 & 1 & 1 \\
loop\_unswitching$^\dag$ & 1 & 1 & 0 & 0 & 1 & 1 & 0 & 0 \\
path\_splitting$^\dag$ & 0 & 0 & 0 & 0 & 1 & 1 & 0 & 0 \\

\bottomrule
\end{tabular}
\begin{tablenotes}
  \small
  \item[$\dag$] benchmarks compiled with loop unrolling and vectorization disabled.
\end{tablenotes}
\end{threeparttable}
\end{table*}

\textbf{Experiments.}
We compile our benchmark with Clang~12 and 18, GCC~9 and 13 with debug symbols, at optimization levels \texttt{O3} and \texttt{Os}, obtaining in total 8 sets of binaries.
We run Microwalk on each binary with a set of 16 randomly generated secret inputs, as recommended by the authors~\cite{WichelmannME18}.
We parse its output, matching for each vulnerable instruction reported in each binary its associated source code line.
\Cref{tab:compilersdiff} reports the number of vulnerable source code lines found in each binary, for each compiler/optimization combination.
Binaries are grouped by their initial CT status, with the top row benchmarks having non-CT source code while the bottom rows are CT.

\textbf{Non-CT benchmarks.} 
As expected, Microwalk reports constant-time violations for the top rows, though the exact numbers differ from one compiler version to another, depending on the benchmarks.
For asymmetric primitives we see the same numbers, while for more complex asymmetric cryptography we tend to see a slight increase in the number of constant-time violations found.
In particular, we note that Microwalk reports 6 additional violations for Clang~18 at \texttt{O3} and 11 for GCC~13 at \texttt{Os}.
For other combinations, the differences are more subtle as some benchmarks see an increase while others see a \textit{decrease}.

\textbf{CT benchmarks.}
For the benchmarks on the bottom rows, Microwalk also reports constant-time violations.
As these benchmarks are compiled from CT source code, these violations must be compiler-introduced.
We manually confirmed that these are real constant-time violations, and not false positives.
As expected we see such violations in our toy examples, already known to be problematic~\cite{GCCPathSplittingBug,GCCJumpThreadingBug}.
GCC introduces constant-time violations in our \texttt{jump\_threading}, \texttt{loop\_unswitching} and \texttt{path\_splitting} benchmarks at \texttt{O3}.
As loop unswitching and path splitting may increase code size, they are disabled in \texttt{Os}, hence why we then see violations only in \texttt{jump\_threading}. A similar scenario plays out for \texttt{loop\_unswitching} in LLVM.
For both GCC and LLVM, we see no constant-time violations in \texttt{ct\_select} functions, despite some being demonstrated in~\cite{SimonCA18}.
Such violations only occur in the i386 back end used by the authors, as it does not support conditional move instructions.
As Microwalk only supports 64-bit binaries, compilers can use such instructions, which is the case in our \texttt{ct\_select} benchmark.
Beyond our toy examples, we see that LLVM~18 adds constant-time violations in our RSA-bearssl and ECDSA-bearssl benchmarks.
Both occur in similar snippets performing a constant-time access into a lookup table used to optimize modular exponentiation and scalar multiplication respectively.
Such violations could leak the index used to access these tables, potentially allowing attacks similar to~\cite{BrumleyH09,InciGI16}.

\textbf{Increase in LLVM.}
We note significant differences in the number of violations reported between LLVM versions. In particular, the previous two found in BearSSL as well the one from \texttt{poly\_frommsg} are not present in LLVM~12.
This is less the case with GCC, where constant-time violations found in one version are often found in another.
These results mirror those of~\cite{SchneiderLP24}, however we must nuance that increase somewhat, as the metrics considered are different.
Indeed, while the authors compared the percentage of binaries containing \textit{at least one} leak, we are comparing the \textit{number} of leaks found in binaries directly.
As such, while we confirm that newer versions of LLVM tend to produce more non-CT binaries, in total only a few constant-time violations are actually added.

\section{Analysis}\label{sec:analysis}

To answer \textbf{RQ2}, we now focus on the compiler-introduced constant-time violations we detected in the previous section.
We isolate the functions responsible for the violations and input them into an online tool, Compiler Explorer~\cite{CompilerExplorer}, which includes the \textit{Opt Pipeline} and \textit{GCC Tree/RTL} viewer tools.
This allows us to finely explore the impact of each optimization pass, as well as their interactions with each other, and determine precisely how they break constant-time guarantees for LLVM and GCC respectively.

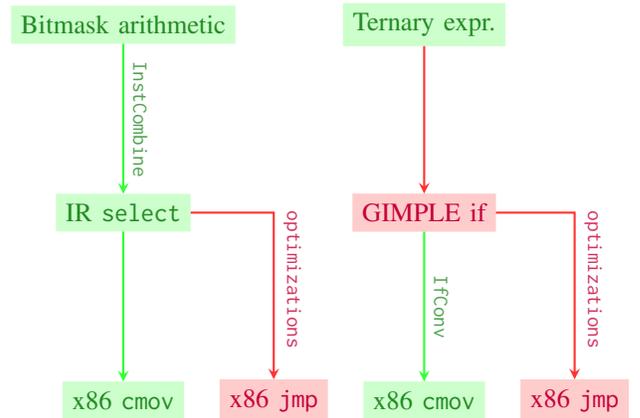
\begin{figure}[t]
  \centering
  \begin{tikzpicture}

    \node[rectangle, thick, draw=green, fill=green, opacity=.2, text opacity = 1, text=ForestGreen] (CTpattern) at (1,5) {Bitmask arithmetic};
    \node[rectangle, thick, draw=green, fill=green, opacity=.2, text opacity = 1, text=ForestGreen] (select) at (1,2.5) {IR \texttt{select}};
    \node[rectangle, thick, draw=green, fill=green, opacity=.2, text opacity = 1, text=ForestGreen] (cmov) at (1,0) {x86 \texttt{cmov}};
    \node[rectangle, thick, draw=red, fill=red, opacity=.2, text opacity = 1, text=BrickRed] (llvmbranch) at (3,0) {x86 \texttt{jmp}};

    \draw[thick, draw=green, fill=green, opacity=.8] (CTpattern) edge["InstCombine" {sloped,text=ForestGreen, font=\footnotesize\ttfamily},-stealth] (select);
    \draw[thick, draw=green, fill=green, opacity=.8] (select) edge[-stealth] (cmov);
    \draw[thick, draw=red, -stealth, opacity=.8] (select) -| (llvmbranch) node[above, sloped, pos=.70, text=BrickRed, font=\footnotesize\ttfamily] {optimizations};

    \node[rectangle, thick, draw=green, fill=green, opacity=.2, text opacity = 1, text=ForestGreen] (CTpattern) at (5,5) {Ternary expr.};
    \node[rectangle, thick, draw=red, fill=red, opacity=.2, text opacity = 1, text=BrickRed] (gimple) at (5,2.5) {GIMPLE if};
    \node[rectangle, thick, draw=green, fill=green, opacity=.2, text opacity = 1, text=ForestGreen] (cmov) at (5,0) {x86 \texttt{cmov}};
    \node[rectangle, thick, draw=red, fill=red, opacity=.2, text opacity = 1, text=BrickRed] (gccbranch) at (7,0) {x86 \texttt{jmp}};

    \draw[thick, draw=red, fill=red, opacity=.8] (CTpattern) edge[-stealth] (gimple);
    \draw[thick, draw=green, fill=green, opacity=.8] (gimple) edge["IfConv" {sloped,text=ForestGreen, font=\footnotesize\ttfamily},-stealth] (cmov);
    \draw[thick, draw=red, -stealth, opacity=.8] (gimple) -| (gccbranch) node[above, sloped, pos=.70, text=BrickRed, font=\footnotesize\ttfamily] {optimizations};

  \end{tikzpicture}
  \caption{Pathways to breaking CT in LLVM (left) and GCC (right).}\label{fig:ctbreakage}
\end{figure}


In general, we found that the main way compilers introduce constant-time violations is by failing to produce a conditional move instruction where possible, instead falling back to a conditional branch.
However, this issue manifests itself differently in both compilers, as shown in~\Cref{fig:ctbreakage}.
In LLVM, such behaviors had been noticed in particular in the i386 backend, where instructions like \texttt{cmov} are unavailable~\cite{SimonCA18}, and more recently in the x86-64 backend under certain conditions~\cite{SchneiderLP24,ZhangB25}.
To our knowledge, we are the first to properly document this behavior in GCC.

\subsection{LLVM}


The Clang front-end and LLVM middle-end readily represent conditional moves using a \texttt{select} IR instruction.
In fact, LLVM has many opportunities to issue this instruction, as its powerful arithmetic optimization engine tends to turn common CT programming patterns to conditional moves.
These optimizations are implemented in the \texttt{InstCombine} pass, and it was noted by~\cite{SchneiderLP24} that improvements in this pass were in large part responsible for the increase in CT violations the authors had found in recent versions of LLVM.
While we confirm that successive applications of \texttt{InstCombine} \textit{undo} bit-mask arithmetic, this \textit{by itself}, does not break constant-time.
Indeed this depends entirely on whether the \texttt{select} IR instruction is kept by further optimizations and whether the x86-64 back-end lowers it into a constant-time \texttt{cmov}.
In the rest of this section, we show that this is not always the case, highlighting the need for our fine-grain analysis.

\begin{figure}[t]

  \begin{subfigure}{\linewidth}
    \begin{lstlisting}
for (u = 1; u < ((uint32_t)1 << k); u++) {
  uint32_t mask;

  mask = -EQ(u, bits);
  for (v = 1; v < mwlen; v++) {
    t2[v] |= mask & base[v];    // CT array access
  }
  base += mwlen;
}
    \end{lstlisting}
    \caption{C source code.} 
    \label{fig:rsa-bearssl-c}
  \end{subfigure}

  \begin{subfigure}{\linewidth}
    \lstset{style=llvmir}
    \begin{lstlisting}
outer_body:
  %EQ = icmp eq i32 %bits, %u   ; inlined call to EQ
  ...
inner_body:
  ; loading base[v]
  %base_v = getelementptr i32, ptr %base, i64 %v
  %base_val = load i32, ptr %base_v, align 4
  ; loading t2[v]
  %t2_v = getelementptr i32, ptr %t2, i64 %v
  %t2_val = load i32, ptr %t2_v, align 4
  ; bit-mask arithmetic replaced by select IR
  %masked_val = select i1 %EQ, i32 %base_val, i32 0
  %res = or i32 %t2_val, %masked_val
  store i32 %res, ptr %t2_v, align 4
  %v_inc = add nuw nsw i64 %v, 1
  br label %inner_cond
    \end{lstlisting}
    \caption{LLVM IR after optimization and inlining of \texttt{EQ}.} 
    \label{fig:rsa-bearssl-llvm}
  \end{subfigure}
  
  \caption{Vulnerable snippet in RSA-bearssl.} 
  \label{fig:rsa-bearssl}
\end{figure}

To help us understand how, let us consider two examples of compiler-introduced CT violations we found in the previous section, one in RSA-bearssl and one in ECDSA-bearssl.
The C source of the former is presented in \Cref{fig:rsa-bearssl-c}, the latter being very similar.
Both examples amount to performing a constant-time read access in an array.
Figure~\ref{fig:rsa-bearssl-llvm} shows the LLVM IR resulting from arithmetic simplifications and function inlining (edited for brevity).
\texttt{InstCombine} optimizes the \texttt{EQ} function into a compare instruction, which is then inlined into the outer loop's body.
Another application of \texttt{InstCombine} there further optimizes the bit-mask arithmetic in the inner loop into a \texttt{select} IR instruction.
At this point in the optimization pipeline, the IR remains constant-time, and the x86-64 would lower \texttt{select} into \texttt{cmov}. 

We now focus on how further optimizations affect this IR snippet.
We use Compiler Explorer to compile this snippet with Clang 18 at \texttt{O3} with various optimizations disabled, and note whether a secret-dependent branch is generated.
For both snippets, \Cref{fig:llvm-details} shows whether it was CT or not depending on the set of optimizations enabled.

\begin{table}[t]
  \caption{Clang optimizations interacting to break CT. Boxes highlight optimizations responsible for adding secret-dependent branches.}
  \label{fig:llvm-details}
  \small
  \begin{subfigure}{\linewidth}
  \resizebox{\linewidth}{!}{
    \begin{tabular}{l ccccccc}
      \toprule
      \textbf{Optimization} & \multicolumn{7}{l}{\textbf{Activated?}} \\
      \midrule
      Loop unswitching  & \cfbox{BrickRed}{\cmark} &        &        &        &        &        &         \\
      Loop unrolling    & \cmark & \cmark &        &        & \cmark & \cmark &         \\
      Loop vectorization& \cmark & \cmark & \cfbox{BrickRed}{\cmark} &        &        &        &         \\
      \texttt{cmov} conversion   & \cmark & \cmark & \cmark & \cfbox{BrickRed}{\cmark} & \cmark &        &         \\
      \midrule
      \textbf{CT?}      & \rxmark & \gcmark & \rxmark & \rxmark & \gcmark & \gcmark & \gcmark \\
      \bottomrule
    \end{tabular}
   }
    \caption{ECDSA-bearssl}
  \end{subfigure}
  \\
  \begin{subfigure}{\linewidth}
  \resizebox{\linewidth}{!}{
    \begin{tabular}{l ccccccc}
      \toprule
      \textbf{Optimization} & \multicolumn{7}{l}{\textbf{Activated?}} \\
      \midrule
      Loop unswitching  & \cfbox{BrickRed}{\cmark} &        &        &        &        &        &         \\
      Loop unrolling    & \cmark & \cmark &        &        & \cmark & \cmark &         \\
      Loop vectorization& \cmark & \cfbox{BrickRed}{\cmark} & \cfbox{BrickRed}{\cmark} &        &        &        &         \\
      \texttt{cmov} conversion   & \cmark & \cmark & \cmark & \cfbox{BrickRed}{\cmark} & \cfbox{BrickRed}{\cmark} &        &         \\
      \midrule
      \textbf{CT?}      & \rxmark & \rxmark & \rxmark & \rxmark & \rxmark & \gcmark & \gcmark \\
      \bottomrule
    \end{tabular}
   }
    \caption{RSA-bearssl}
  \end{subfigure} 
\end{table}

\textbf{Loop Unswitching.}
Loop unswitching usually applies in branches within loops. However, as implemented in LLVM, \texttt{SimpleLoopUnswitchPass} considers \texttt{select} IR instructions as well.
Since the \texttt{select} IR instruction in the inner loop is independent of its index, loop unswitching applies, turning it into a conditional branch brought in the outer loop.
The inner loop is duplicated in both sides of the branch: one side where \texttt{t2[v]} is updated, and one side with no effects. This latter side is deleted by further passes, leaving us with a secret-dependent branch skipping the inner loop.
We see this behavior in both ECDSA-bearssl and RSA-bearssl.

\textbf{Loop Vectorization.}
Disabling loop unswitching preserves the \texttt{select} IR instruction, but that is not enough to obtain constant-time assembly.
Indeed, loop vectorization is applied later in the pipeline. This has the effect of duplicating the inner loop into two versions: one vectorized with a factor of 4 and one serving as the scalar epilogue.
In the vectorized loop, the back-end lowers \texttt{select} IR instructions into a branch, as no vector \texttt{cmov} instructions exist in x86 (although a similar effect can be achieved using a \texttt{pblend} instruction).

We see this behavior in RSA-bearssl, but not in ECDSA-bearssl. In the latter, the inner loop is unrolled and basic block vectorization is then applied.
In this case, the \texttt{select} IR instructions are lowered back into vectorized \texttt{and} instructions, recreating the original constant-time bit-mask arithmetic.
However, if we disable loop unrolling, the inner loop is vectorized just as in RSA-bearssl and a constant-time violation is introduced. 
This shows how subtle interactions between optimization passes can break constant-time guarantees but also, sometimes, \textit{restore} them.

\textbf{\texttt{cmov} conversion.}
Disabling loop vectorization does not solve the issue completely as the scalar epilogue loop is not safe from optimizations by the back-end either.
Indeed, while its \texttt{select} IR instruction is initially lowered into a \texttt{cmov}, the back-end converts it into a conditional branch.
As this can potentially save a memory load, using a branch there, even secret-dependent, is deemed profitable by the compiler.

After disabling this conversion and the above passes, we finally obtain constant-time binary for RSA-bearssl. \texttt{cmov} conversion is also responsible for the constant-time violation in the \texttt{poly\_frommsg} benchmark from \Cref{tab:compilersdiff}.

\subsection{GCC}

In contrast to LLVM, GGC does not internally represent conditional moves so easily.
Instead, they are represented as regular conditional branches by default~\cite[page 227]{GCCinternals}.
Despite this, as we saw in the previous section, GCC preserves constant-time more often. This can be explained by two factors.
First, GCC is much more conservative with its simplifications, meaning bit-mask arithmetic rarely gets optimized into a branch.
Second, GCC includes an \textit{if-conversion} pass that transforms branches into branchless code where possible, potentially \textit{eliminating} secret-dependent branches.
As such, typical constant-time programming patterns tend to be preserved, or at worse, get lowered into \texttt{cmov} instructions.

In our benchmarks, constant-time violations are rarely added by GCC. The only exception is code using the ternary conditional operator \texttt{a ? b : c}.
This operator is internally treated as a branch by GCC, though with \textit{if-conversion}, constant-time should be preserved.
However we note it is possible for optimizations to transform IR branches in such a way where the conversion cannot apply.
These then get lowered into conditional jumps by the back-end, introducing a constant-time violation.

We see such behaviors in successive ternary conditionals with overlapping conditions. Jump threading then applies and merges the conditionals, removing an opportunity to lower them to \texttt{cmov}.
Similar behaviors occurs in ternary conditionals within loops, as path splitting and loop unswitching can apply.

\section{Mitigations}\label{sec:evaluation}
We saw in the previous section how compiler optimizations can interact to break source code level constant-time guarantees.
Our benchmark allowed us to identify precisely a set of problematic optimizations in both LLVM and GCC.
In this section, to answer \textbf{RQ3} we first determine if disabling these optimizations is enough to mitigate compiler-introduced constant-time violations.
We then evaluate the performance impact of disabling such optimizations.

It is important to remember that \textbf{our approach does not aim to fix code that is non-constant time prior to compilation, as no compiler configuration can retroactively make these implementations CT}.

We disable problematic optimizations using the option flags in both LLVM and GCC. The exact set to disable depends on the compiler.

\textbf{LLVM flags.} We disable \texttt{cmov} conversion, loop unswitching and loop vectorization.
Loop vectorization is disabled using a flag already included in Clang.
For the others, we need to pass undocumented options to the LLVM back-end using \texttt{-mllvm}.
\texttt{cmov} conversion is disabled in both its x86 back-end pass and in the middle-end pass \texttt{CodeGenPrepare}.
Loop unswitching is disabled by setting a threshold rendering this optimization always unprofitable:
\begin{itemize}
  \item \texttt{cmov} conversion: \texttt{-mllvm \ddash x86-cmov-converter=false -mllvm \ddash disable-cgp-select2branch=true}
  \item Loop unswitching: \texttt{-mllvm \ddash unswitch-threshold=1}
  \item Loop vectorization: \texttt{-fno-vectorize}
\end{itemize}

\textbf{GCC flags.} We disable loop unswitching, jump threading and path splitting. For all these, flags are already available within GCC.
\begin{itemize}
  \item Loop unswitching: \texttt{-fno-unswitch-loops}
  \item Jump threading: \texttt{-fno-thread-jumps}
  \item Path splitting: \texttt{-fno-split-paths}
\end{itemize}

\subsection{Preserving constant-time}\label{sec:evaluation:CT}

\CatchFileDef{\tablemitigationO3}{table_mitigationO3.tex}{}
\begin{table*}[ht]
\centering
\caption{Number of vulnerable lines per binaries, comparison between our mitigations  and a baseline without mitigations for LLVM~18 and GCC~13 at \texttt{O3}. We highlight increases in number of vulnerabilities with \inc{} and decreases with \dec. The first group of binaires is produced from non-CT source code, and the second one from CT source code.}
\label{tab:mitigationO3}
\small
\begin{threeparttable}
\begin{tabular}{l|ccc|cc}
  \toprule
     & \multicolumn{3}{c}{LLVM 18} & \multicolumn{2}{c}{GCC 13} \\
     Binaries & No mitig. & Mitig.+Vect. & Mitig. & No mitig. & Mitig. \\

  \midrule

  \tablemitigationO3

\bottomrule
\end{tabular}
\end{threeparttable}
\end{table*}

\textbf{Experimental setup.} We recompile the benchmark used in Section~\ref{sec:dataset} using Clang~18 and GCC~13, with our mitigating set of flags.
As disabling vectorization might have a significant impact on performance, we also compile a variant where vectorization is kept enabled for Clang.
This allows us to properly gauge if the side-channel protection added by disabling vectorization is worth its performance impact.
We run Microwalk on the generated binaries, following the same setup as in Section~\ref{sec:setup}.

\textbf{Results.} Table~\ref{tab:mitigationO3} compares the number of vulnerable source code lines reported by Microwalk between binaries compiled with and without our mitigations (``No mitig.'' and ``Mitig.'' respectively).
In the case of LLVM, column ``Mitig.+Vect.'' reports results obtained with mitigations but with loop vectorization kept enabled.
For both LLVM and GCC, we are able to remove all the constant-time violations the compilers we studied added in CT source code.
Disabling loop vectorization is necessary in the case of the RSA-bearssl benchmark, for LLVM specifically.
In non-CT source code, we tend to reduce the number of constant-time violations for LLVM.
In fact, in the case of RSA-mbedtls, the number of constant-time violations found with our full mitigations is equivalent to what we saw with Clang~12 in Section~\ref{sec:dataset}.
We conclude that our set of option flags is able to reduce the number of compiler-introduced constant-time violations, thus answering the first part of of \textbf{RQ3}.

\subsection{Performance impact}\label{sec:evaluation:perf}

\CatchFileDef{\tableperformanceO3}{table_performanceO3.tex}{}
\begin{table}[ht]
\centering
\caption{Performance impact on MbedTLS and BearSSL as a percentage difference from a baseline without mitigations. The metric used is the average number of operations per second, \ie negative numbers indicate a performance \textit{decrease}.}
\label{tab:mitigationO3perf}
\small
\resizebox{\linewidth}{!}{
\begin{threeparttable}
\begin{tabular}{l|cc|c}
  \toprule
     & \multicolumn{2}{c}{LLVM 18} & \multicolumn{1}{c}{GCC 13} \\
     Benchmark & Mitig.+Vect. & Mitig. & Mitig. \\

  \midrule

  \tableperformanceO3

\bottomrule
\end{tabular}
\end{threeparttable}
}
\end{table}

To evaluate the performance impact of our mitigating set of compilation flags, we reuse the benchmark provided within BearSSL and MbedTLS.
These execute cryptographic operations from their respective libraries repeatedly, reporting for each operation the average amount of input data processed per second.
We compare the results of these benchmarks when the libraries are compiled with our mitigations, compared to a baseline of the default \texttt{O3} optimization level. For LLVM 18 we include two variants, one with all mitigations and one keeping vectorization enabled. For GCC 13, we keep vectorization enabled.
We use the same laptop as in Section~\ref{sec:setup}, with one physical core isolated from the Linux scheduler using \texttt{isolcpus}.
Experiments are realized on that core, with low system load on other cores.

Table~\ref{tab:mitigationO3perf} reports the percentage difference with the baseline. For brevity, we only report benchmarks roughly corresponding to those Table~\ref{tab:compilersdiff}, as well as the mean and standard deviations for BearSSL and MbedTLS in general. The full table is available in the artifact. 

\textbf{Results.} For LLVM 18, we found that with our mitigations BearSSL had on average a 0.80\% ($SD=4.98$) performance decrease with vectorization, and a 3.30\% ($SD=8.22$) decrease without.
For MbedTLS, we saw a slight decrease of 0.77\% ($SD=2.21$) and 0.71\% ($SD=0.71$), with and with vectorization respectively.
The trend is similar for GCC 13, where we saw a slight performance decrease of 0.43\% ($SD=2.63$) and 1.14\% ($SD=3.86$) for BearSSL and MbedTLS respectively.
Overall, the trend seems to indicate a slight performance decrease, however we note that with large standard deviations, this decrease is not significant. 
We conclude that the performance impact of our mitigating set of flags is negligible.

\section{Discussion and Related Work}\label{sec:discussion}
In this section, we examine related work in light of our findings, highlighting how our approach compares to existing tools and compiler-level defenses. Building on this analysis, we outline practical recommendations to improve the robustness of constant-time implementations. We conclude with directions for future research.

\subsection{Detecting and correcting CT violations} 
Constant-time violations in software can enable side-channel attacks. 
These issues are often addressed through CT programming, which offers rapid and low-overhead mitigation strategies~\cite{LouZJ21}. 
As a result, numerous detection tools have been developed over the years. However, their adoption by developers remains limited. 
Jancar \etal\cite{JancarFB22} and Fourné \etal\cite{FourneBJSSBFA24} attribute this low adoption in part to poor usability. 
In parallel, Geimer \etal\cite{GeimerVR23} examined the technical limitations that current tools face when attempting to detect more complex vulnerabilities.
These tools rely on both static~\cite{DanielBR20} and dynamic analysis~\cite{WichelmannME18}, and operate at various levels of abstraction, including source code~\cite{ChattopadhyayR18}, LLVM IR~\cite{disselkoen2020finding}, and binary code~\cite{DanielBR20,WichelmannME18}. Subsequent work has aimed at automatically correcting constant-time violations, either by linearizing control-flow and data-flow constructs~\cite{BorrelloDQ21}, or by replacing unsafe instructions with constant-time equivalents~\cite{DineshGF22,FlandersSMGK24}.
While our work relies on existing detection tools, it does not aim to improve them or promote their adoption.

\subsection{Compiler-introduced CT violations}
Despite growing awareness that compilers can undermine constant-time guarantees, the issue has long remained under the radar, with few studies offering an analysis of its prevalence, impact, and root causes.
The following works are the closest to our work.
The seminal work from Simon \etal\cite{SimonCA18} and Daniel \etal\cite{DanielBR20} provided early insights, but was limited to a small number of illustrative code snippets.
More recently, Schneider \etal\cite{SchneiderLP24} and Gerlach \etal\cite{GerlachPS25}, in work concurrent with ours, conducted extensive quantitative analyses of compiler-induced side-channel leakages, shedding light on their prevalence and impact.
Thanks to a custom analysis pipeline, the former are able to detect leakages in different combinations of optimization levels, architecture and compiler versions.
However, their investigation on how compilers passes introduce these leakages remains limited, and covers only LLVM.
Gerlach \etal\cite{GerlachPS25} similarly design an analysis pipeline based on a combination of detection tools applying it to numerous compilers however, unlike our approach, they do not investigate the root causes of these leakages.
Neither propose mitigation strategies.

Orthogonal to our work, Xu \etal\cite{XULD23} conducted a broader study on compiler-introduced bugs. 
While their analysis covers a wide range of issues, it pays only limited attention to constant-time violations--illustrated by a few examples, in part due to challenge C1--and does not analyze compiler optimizations.
Finally, Arranz Olmos \etal\cite{abs-2501-04183} examined a complementary problem: the preservation of constant-time properties in decompilers. Their findings suggest that analysis tools lifting binary programs to intermediate representations may be unsound if the decompiler is not transparent--i.e., if it introduces or removes constant-time violations.

\subsection{Principled but underused: Secure compilation} 
A principled approach to addressing the issues outlined above is to ensure that compilers themselves preserve constant-time properties throughout the compilation process~\cite{BartheGL18}.
Notably, Barthe \etal\cite{BartheBG20} extended the formally verified CompCert compiler to provide constant-time preservation guarantees, and proved its correctness in this regard.
However, such secure compilers remain rarely used in practice, as they offer limited architecture support and performance compared to mainstream compilers such as GCC and LLVM.
An alternative line of work has explored patching commodity compilers--e.g., by adding dedicated functions~\cite{ctchoose}--to enforce constant-time properties.
Yet these efforts have not been merged upstream and are no longer maintained, rendering them unusable.

	\subsection{Our recommendations}
The bleak picture of side-channel security today can be traced back to the overall incentive driving the industry: maximizing performance~\cite{Pornin25}.
Compiler developers are incentivized to improve performance by adding optimizations that radically transform the original source code, sometimes interfering with developers' intents.
On the other hand, developers try to improve \textit{their} performance by minimizing the time spent implementing features, and minimize the risks of adding bugs.
Compiler developers too are subject to this sort of pressure.
It is thus unsurprising that principled approaches tend to be underused and that popular compilers are reluctant to add such guarantees~\cite{LLVMdev19,GCCMailCmov}: the overall incentive is simply \textit{not there}.

While side-channel secure compilers and languages offer strong theoretical guarantees and should be considered where feasible, they remain difficult to integrate into existing workflows. As such, in the absence of widespread adoption, developers may instead rely on the set of compiler flags we introduced in \Cref{sec:evaluation} to mitigate leakage in practice.
This reflects a broader reality: bridging the gap between source-level guarantees and compiler behavior is not just a technical challenge, but also a matter of ecosystem inertia.
In the absence of structural change, practical mitigations such as ours offer a realistic path to reducing the impact of compiler-induced vulnerabilities.

\subsection{Limitations and Future work}
While our approach provides a clearer understanding of how compilers introduce constant-time violations, it comes with some limitations. 
In particular, the analysis requires a substantial amount of manual effort to identify problematic optimization passes, and does not guarantee completeness: there may exist other passes or interactions that introduce vulnerabilities but were not observed in our current benchmarks.
That said, our empirical evaluation shows that disabling the small set of identified passes successfully eliminates \emph{all} compiler-introduced violations in our set of known constant-time implementations. This suggests that we have likely uncovered the most impactful passes in practice.
To further improve coverage and reduce manual effort, a promising direction would be to develop a dedicated detection tool operating at the compiler IR level, and run it between each pass. 
By comparing the number of constant-time violations at each stage, such a tool could help automatically pinpoint which passes introduce new leaks.

In work concurrent to ours, Zhang \etal\cite{ZhangB25} introduced a CT analysis tool based on LLVM IR, which can be added as a plugin to a regular LLVM compilation pipeline.
This ease-of-use makes it a prime candidate for an analysis of LLVM passes, however at the time of writing, this tool has yet to be open-sourced.
Another limitation is that it can only analyze middle-end passes, but as we saw in \Cref{sec:analysis}, the back-end is sometimes responsible for introducing leakages by lowering \texttt{select} IR instructions to branches.
While the authors included an option to report secret-dependent \texttt{select} as leakages, it would still require us to manually check the binary to see how it was lowered.
Similarly, Pitchfork~\cite{disselkoen2020finding} can find constant-time violations in LLVM IR, but the tool has gone unmaintained and does not support versions after LLVM 12. 
To our knowledge, no equivalent of such tools exist for GCC.

\section{Conclusion}\label{sec:conclusion}

This paper presents a fine-grained analysis of how mainstream compilers such as GCC and LLVM break constant-time guarantees.  
By attributing violations to specific optimization passes and studying their interactions, we go beyond existing work that focuses only on compiler versions or optimization levels.
We proposed a simple and effective mitigation strategy: disabling a targeted set of passes using both standard and undocumented compiler flags.  
This approach removes all compiler-introduced vulnerabilities in known constant-time implementations from our benchmarks, with low performance overhead and no change to the source code.

Unlike solutions that require rewriting cryptographic code in a dedicated language or relying on formally verified compilers, our method is immediately deployable and fits within current developer workflows.  
While such principled approaches remain desirable in theory, their adoption remains limited in practice.

Our results highlight the value of lightweight, actionable defenses that strengthen security without requiring heavy engineering or social adoption efforts.  
A key challenge for future work is to further bridge the gap between verified compilation guarantees and the complexity of real-world toolchains, potentially by combining formal analysis with scalable, pass-level detection techniques.

\ifAnon
\else
\section*{Acknowledgments}
This work benefited from the support of the AID and ANR-19-CE39-0007 MIAOUS.
\fi

\bibliographystyle{plain}
\balance
\bibliography{biblio}

\appendix

\textbf{Impact of unrolling.} \Cref{tab:compilersunroll} shows the impact of loop unrolling on the number of vulnerabilities found in binaries following two metrics: instructions and source code lines. All binaries are compiled with LLVM 18 at \texttt{O3}, with loop unrolling either enabled (\textcolor{ForestGreen}{\cmark}) or disabled (\textcolor{BrickRed}{\xmark}).

We note that disabling loop unrolling appears to have a significant impact on the number vulnerable instructions, in particular for AES-mbedtls benchmarks or ECDSA-bearssl.
However this does not necessarily translate to a more ``secure'' binary. Using source code lines as a metric, we can see that just as many constant-time violations were generated, loop unrolling simply tends to duplicate them.

\textbf{Toy Examples.} \Cref{lst:toy-examples} shows the code for the toy examples we used in our benchmarks. The toy examples for the \texttt{ct\_select} benchmark can be found in~\cite{SimonCA18}. 

\lstset{style=customc}
\begin{lstlisting}[label={lst:toy-examples}, caption={Source code of our toy examples}]
// https://groups.google.com/a/list.nist.gov/g/pqc-forum/c/hqbtIGFKIpU
void poly_frommsg(int16_t r[256], const uint8_t msg[32]) {
  unsigned int i,j;
  int16_t mask;

  for(i=0;i<256/8;i++) {
    for(j=0;j<8;j++) {
      mask = -(int16_t)((msg[i] >> j)&1);
      r[8*i+j] = mask & 1665;
    }
  }
}

void loop_unswitch(char* x, char* y, size_t n, bool w) {
 for (int i = 0; i < n; i++) {
    int t = y[i];
    x[i] += t;
    y[i] = w ? 0 : t;   // secret-dependent cond-move
  }
}

// https://kristerw.github.io/2022/05/24/branchless/
int jump_threading(uint8_t /*secret*/ a) {
  int r = ((a < 10) != 0) << 4;
  r |= ((a > 12) != 0) << 2;
  return r;
}

// https://gcc.gnu.org/bugzilla/show_bug.cgi?id=112402
int path_splitting(int8_t /*secret*/ *p, size_t n) {
  int s = 0;

  for (int i = 0; i < n; i++) {
    int t;
    t = (p[i] >= 0) ? 1 : -1;
    s+=t;
  }

  return s;
}
\end{lstlisting}

\newcommand{\tableunroll}{AES-CBC-bearssl (T) & 36 & 36 & 32 & 32 \\ 
AES-GCM-bearssl (T) & 36 & 36 & 32 & 32 \\ 
AES-CBC-mbedtls (T) & 104 & \textbf{68} \dec & 23 & 23 \\ 
AES-GCM-mbedtls (T) & 108 & \textbf{71} \dec & 26 & 26 \\ 
RSA-mbedtls (PKCS) & 144 & \textbf{141} \dec & 47 & \textbf{48} \inc \\ 
RSA-mbedtls (OAEP) & 143 & 143 & 48 & \textbf{49} \inc \\ 
ECDSA-mbedtls & 147 & \textbf{138} \dec & 64 & \textbf{61} \dec \\ 
\midrule
AES-CBC-bearssl (BS) & 0 & 0 & 0 & 0 \\ 
AES-GCM-bearssl (BS) & 0 & 0 & 0 & 0 \\ 
PolyChacha-bearssl (CT) & 0 & 0 & 0 & 0 \\ 
PolyChacha-bearssl (SSE2) & 0 & 0 & 0 & 0 \\ 
RSA-bearssl (OAEP) & 1 & 1 & 1 & 1 \\ 
ECDSA-bearssl & 7 & \textbf{1} \dec & 1 & 1 \\ 
PolyChacha-mbedtls & 0 & 0 & 0 & 0 \\ 
}
\begin{table}[h!]
\centering
\caption{Number of vulnerable instructions and source code lines per binaries compiled with LLVM 18 at \texttt{O3}, comparison between unrolling enabled (\textcolor{ForestGreen}{\cmark}) and disabled (\textcolor{BrickRed}{\xmark}).}
\label{tab:compilersunroll}
\small
\begin{threeparttable}
\begin{tabular}{l|cccc}
  \toprule
     & \multicolumn{2}{c}{Instr.} & \multicolumn{2}{c}{Lines} \\
  
  Binaries & \textcolor{ForestGreen}{\cmark} & \textcolor{BrickRed}{\xmark} & \textcolor{ForestGreen}{\cmark} & \textcolor{BrickRed}{\xmark} \\

  \midrule

  AES-CBC-bearssl (T) & 36 & 36 & 32 & 32 \\ 
AES-GCM-bearssl (T) & 36 & 36 & 32 & 32 \\ 
AES-CBC-mbedtls (T) & 104 & \textbf{68} \dec & 23 & 23 \\ 
AES-GCM-mbedtls (T) & 108 & \textbf{71} \dec & 26 & 26 \\ 
RSA-mbedtls (PKCS) & 144 & \textbf{141} \dec & 47 & \textbf{48} \inc \\ 
RSA-mbedtls (OAEP) & 143 & 143 & 48 & \textbf{49} \inc \\ 
ECDSA-mbedtls & 147 & \textbf{138} \dec & 64 & \textbf{61} \dec \\ 
\midrule
AES-CBC-bearssl (BS) & 0 & 0 & 0 & 0 \\ 
AES-GCM-bearssl (BS) & 0 & 0 & 0 & 0 \\ 
PolyChacha-bearssl (CT) & 0 & 0 & 0 & 0 \\ 
PolyChacha-bearssl (SSE2) & 0 & 0 & 0 & 0 \\ 
RSA-bearssl (OAEP) & 1 & 1 & 1 & 1 \\ 
ECDSA-bearssl & 7 & \textbf{1} \dec & 1 & 1 \\ 
PolyChacha-mbedtls & 0 & 0 & 0 & 0 \\

\bottomrule
\end{tabular}
\end{threeparttable}
\end{table}

\end{document}